\documentclass{egpubl}
\usepackage{eg2018}

 \JournalSubmission    
 \electronicVersion 

\ifpdf \usepackage[pdftex]{graphicx} \pdfcompresslevel=9
\else \usepackage[dvips]{graphicx} \fi

\PrintedOrElectronic

\usepackage{t1enc,dfadobe}

\usepackage{egweblnk}
\usepackage{cite}

\usepackage{tikz}
\usetikzlibrary{arrows,fadings,patterns,shapes}
\usetikzlibrary{decorations.pathmorphing}
\usetikzlibrary{decorations.pathreplacing}
\usetikzlibrary{calc}
\usepackage{adjustbox}
\usepackage{alphalph}

\usepackage{flafter}

\title[Graphic Narrative with Interactive Stylization Design]%
      {Graphic Narrative with Interactive Stylization Design}

\author[I. Garcia-Dorado, P. Getreuer, M. Le, R. Debreuil, Alex K., P. Milanfar]
{\parbox{\textwidth}{\centering I. Garcia-Dorado and P. Getreuer and M. Le and R. Debreuil and A. Kauffmann and P. Milanfar \\ }
        \\ 
{\parbox{\textwidth}{\centering Google Research
       }
}
}

\newcommand{\units}[1]{{}\hspace*{0.2em}\ensuremath{\mathrm{#1}}}

\newcommand{\mcomment}[1]{\textcolor{red}{[\small{}#1]}}
\newcommand{\todo}[2][]{\mcomment{%
\textsc{Todo}\ifthenelse{\equal{#1}{}}{}{(#1)}: #2}}

\begin{document}
\teaser{
 \includegraphics[trim={0.0cm 8.1cm 0.0cm 0.0cm},clip,width=\linewidth]{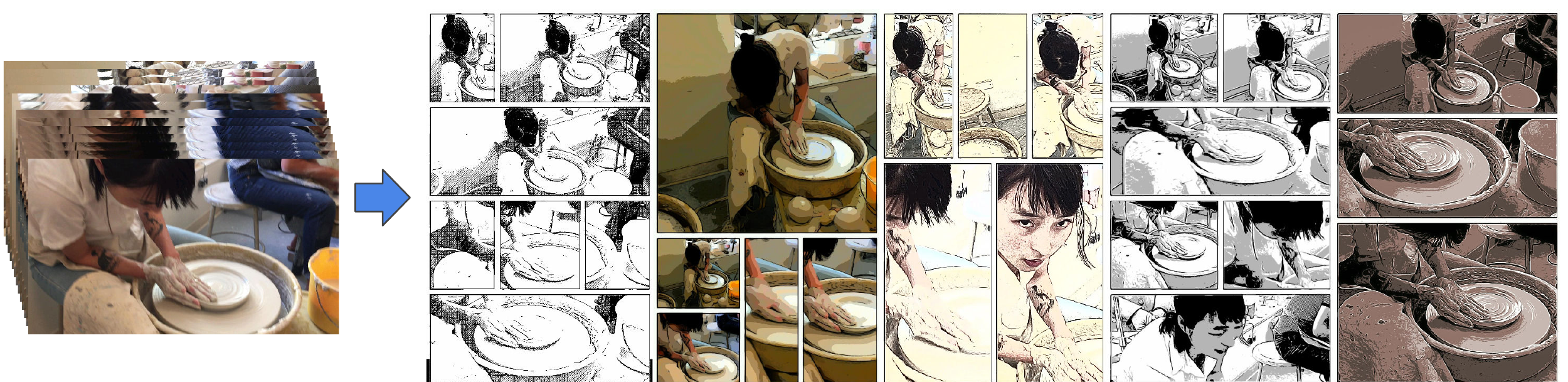}
 \centering
  \caption{Our system converts a set of input images into a series
  of storyboards with varying layouts, images, and stylization.}
\label{fig:teaser}
}

\maketitle
\begin{abstract}

We present a system to convert any set of images (e.g., a video clip or a photo
album) into a storyboard. We aim to create multiple pleasing graphic
representations of the content at interactive rates, so the user can explore and
find the storyboard (images, layout, and stylization) that best suits their
needs and taste. The main challenges of this work are: selecting the content
images, placing them into panels, and applying a stylization. For the latter, we
propose an interactive design tool to create new stylizations using a wide range
of filter blocks. This approach unleashes the creativity by allowing the user to
tune, modify, and intuitively design new sequences of filters. In parallel to
this manual design, we propose a novel procedural approach that automatically
assembles sequences of filters for innovative results. We aim to keep the
algorithm complexity as low as possible such that it can run interactively on a
mobile device. Our results include examples of styles designed using both our
interactive and procedural tools, as well as their final composition into
interesting and appealing storyboards.


\end{abstract}

--------------------------------------------------------------------
\section{Introduction}
A \textit{storyboard} is a series of image panels representing a sequence of
actions that will be performed and/or captured in motion, such as in a film,
an animation, or onstage. Storyboards were first developed at Walt Disney
Studios during the early 1930s~\cite{whitehead2004animation}. The idea was, 
during early stages of brainstorming and development, to draw key scenes on separate
sheets of paper and pin them to create a narrative bulletin board story. This
allowed them to design and iterate faster by interchanging the 
order and by replacing panels with a more interesting ones at any point.
Films have relied on this approach since the days of silent cinema.
Nowadays, storyboards are used to illustrate, summarize,
and pre-visualize many kinds of dynamic content including
TV shows, movies, plays, novels, and even software. Storyboards represent
dynamic content in a concise and explanatory manner, similar in many ways to
comics. The main distinction is while storyboards focus mainly on the action and
aim to convey information, comics focus on the conversation (with speech
bubbles) with the goal of entertainment. 

Our approach uses storyboarding as an intermediate representation between videos and
comics, extracting representative frames and laying them out on a stylized single page.
We do not seek to optimize the layout, the content, and the stylization to
create a unique output. 
Our goal is to allow a user to quickly explore dozens or even hundreds of
alternatives and to select the representation(s) that best fit his or her
requirements (See Fig.~\ref{fig:teaser}). Quick exploration and filtering
require algorithms that run at near real-time speeds on mobile devices,
necessarily limiting their complexity. Moreover, our approach applies beyond video 
to any set of input images such as photo albums and images from
multiple different sources. In summary, our goal is to translate videos and
images into storyboards that present the scenes within them in a
visually pleasing, effective, and interesting manner.

The stylization filters and several other aspects developed here are components of Google Research's "Storyboard" app~\cite{storyboard2017app}. A variant of the described pipeline is used in this application.

Central to the visual impact of our storyboards is stylization. Most recent work
in stylization has focused on convolutional neural network style transfer
algorithms (e.g.,~\cite{gatys2016image}). However, as our results will show, the
resulting effects are hard to control, despite recent efforts to make them more
tunable~\cite{gatys2016controlling}. Alternatives such as abstraction and 
filtering focus on creating one appealing stylization that would satisfy the
requirements (e.g., abstraction simplifies an image by removing details). Our
goal is to create a filter block framework that comprises both simple and
advanced filters that enable creative control and fine tuning. We will also
explore a procedural approach, inspired by~\cite{muller2006procedural}, for
generating novel filters.

The main contributions of this work can be summarized as follows:
\begin{itemize}
\item We propose an automatic system to convert an album or video into storyboards, i.e., interesting and visually pleasing visual stories. Our method focuses on speed so it can interactively run on a mobile device. 
\item We present an interactive design tool for stylization. This tool allows tuning, modifying, and designing styles on the fly, unleashing creativity.
\item We propose a novel procedural style generation technique built on our design tool. Using simple rules we create original stylization effects with a small number of filter blocks.
\end{itemize}

The rest of the paper is organized as follows: Section~\ref{sec:previos_work} reviews previous work. Section~\ref{sec:overview} is an overview of our system. Section~\ref{sec:image_selection} describes the basic process for discarding blurry and near-duplicate images. Section~\ref{sec:framing} explains the process for selecting and framing input images. Section~\ref{sec:stylization} introduces our design application and the filter blocks. Section~\ref{sec:results} presents our results. Finally, Section~\ref{sec:conclusions} contains our conclusions and future work.

\section{Previous Work}\label{sec:previos_work}

We review work in comic generation, summarization, re-targeting, and stylization.

\subsection{Movie to Comic Generation}

The papers most related to our work are those focused on generating a comic from a movie sequence.

Ryu et. al~\cite{ryu2008cinetoon} present a semi-automatic method for converting a film into a comic. The user manually selects the frames, relevant stickers, and word balloons. Then, this is integrate into a comic using a black and white cartoonization effect.
Chu et al.~\cite{chu2015optimized} propose a method to transform a temporal image sequence into a comic-based presentation using a genetic algorithm. This work focuses on the page allocation, layout, and speech balloon placement.
Wang et al.~\cite{wang2012movie2comics} present a method for automatically generating a comic representation of a movie. The aim of the work is to keep all the information (including the script). Thus, this work focuses on finding the shots where a character talks so they can draw the appropriate speech bubble. As a result, a twenty-minute video becomes a seventy-page comic. 
Jing et al.~\cite{jing2015content} extend this approach by proposing a content-driven layout system that generates layouts dynamically rather than relying on pre-defined templates.

\noindent In contrast to all these methods, our work: 
\begin{itemize}
\item allows as input any set of images (i.e., it is not limited to videos),
\item does not require any additional information such as scripts,
\item eschews summarization in favor of exploration, and
\item focuses on interactivity and responsiveness.
\end{itemize}
The only significant overlap is the cartoon stylization step where we present a novel interactive tool for designing styles.

\subsection{Video Summarization}
Summarization techniques focus on selecting good individual frames or shots from a video sequence to create a summary.

Keyframe-based approaches attempt to extract the different interesting parts of the video using low-level features (e.g., image difference~\cite{zhang1997integrated}) or higher-order descriptions of the video (e.g., creating hierarchical visual models\cite{liu2010hierarchical}). Saliency detection~\cite{liu2011learning} methods rely on finding interesting objects and using them to identify important frames of the video.

More recent work uses supervised deep learning to perform video summarization. Gygli et al.~\cite{gygli2015video} present an approach to learn the importance of global characteristics optimized for multiple summarization objectives.
Zhang et al.~\cite{zhang2016summary} use training videos to transfer summary structures non-parametrically.

All these techniques look for a uniquely representative sub-sample of shots to summarize the video. While our work might benefit from these techniques, we require a simple and quick process because our objective is different: given an input, produce a wide range of options for a user to select among.


\begin{figure*}[tb]
\centering
\includegraphics[width=\linewidth]{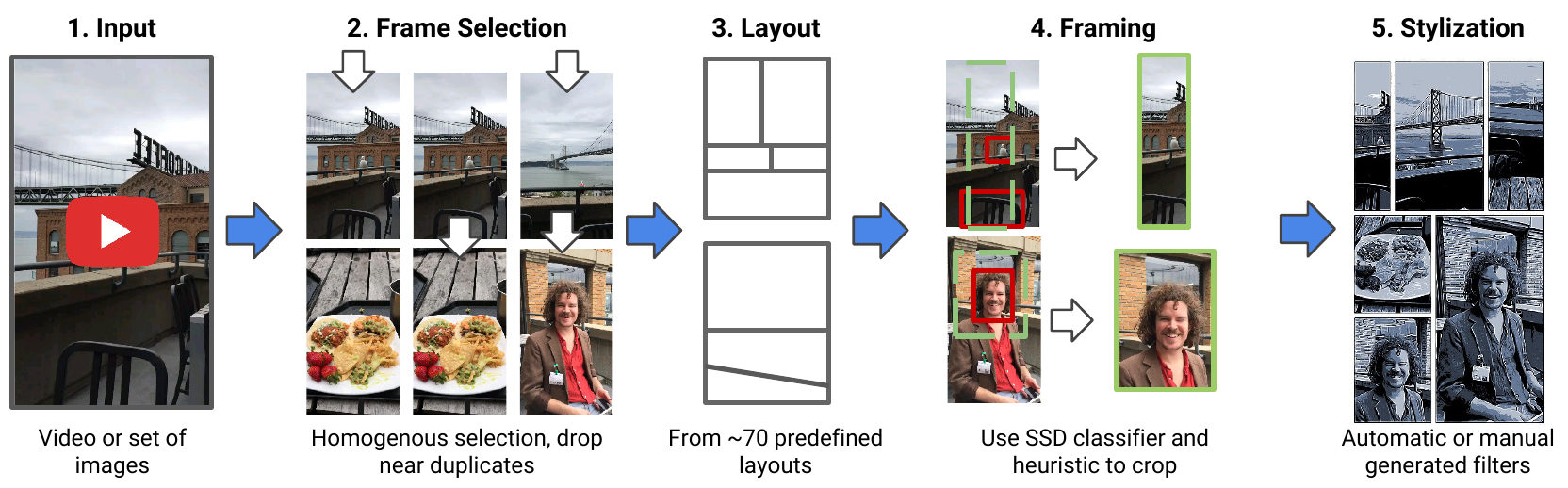}
\caption{\label{fig:pipeline}Pipeline.}
\end{figure*}

\subsection{Stylization}\label{subsec:stylization}

Stylization of images and videos represents a broad research area. We review three main groups: filtering, video stylization, and style transfer.

\begin{itemize}
\item
\textbf{Filtering and Abstraction.} 


Winnem{\"o}ller et al.~\cite{winnemoller2012xdog} explore Difference-of-Gaussians (XDoG) to create interesting sketch and hatching effects. Kang et al.~\cite{kang2007coherent} extend XDoG by adding an edge tangent flow block to create smooth edges.

Kyprianidis and D{\"o}llner~\cite{kyprianidis2008image} use oriented separable filters and XDoG to achieve a high level of image abstraction. Kang et al.~\cite{kang2009flow} improve the level of abstraction from this approach by adding a flow-based step. Other more complex algorithms simplify images using an advanced multi-scale detail image decomposition~\cite{talebi2016fast}.

For a more comprehensive survey of artistic stylizations we refer the reader to Kyprianidis et al.~\cite{kyprianidis2013state}.

We use some of these filters as building blocks for our stylization blocks as we explain in Section~\ref{sec:stylization}.

\item
\textbf{Video Stylization.}

As an extension of image filtering, some works have focused on speeding up the stylization process to run at interactive rates. Winnem{\"o}ller et al.~\cite{winnemoller2006real} use a bilateral filter iteratively to abstract the input, then quantize the background color and overlay XDoG to produce strong outlines. Our system can achieve similar results using a different set of filters. Barnes et al.~\cite{barnes2015patchtable} precompute a multidimensional hash table to accelerate the process of finding replacement patches. This structure enables them to stylize a video in real time using a large collection of patch examples. 

\item
\textbf{Style Transfer.} 

As an alternative to explicit filter creation, a wide range of works have developed a technique called \textit{style-transfer}. Style transfer is the process of migrating a style from a given image (reference) to the content of another (target), i.e., synthesizing a new image which is an aesthetic mixture of the two. Recent work on this problem uses Convolutional Neural Networks (CNN). Gatys et. al.~\cite{gatys2016image} pose the style-transfer problem as an energy minimization task, seeking an image close to the target using a CNN as score. Elad and Milanfar~\cite{elad2017style} present an approach based on texture synthesis. Their approach copies patches from the reference image to the target while maintaining the main features of the content image using a hierarchical structure.

Despite the visual appeal of these approaches, their complexity is a major drawback. The method described by Gatys et. al.~\cite{gatys2016controlling} can take up to an hour to compute a single image. More recent works have focused on addressing this issue. Johnson et al.~\cite{johnson2016perceptual}, achieve real-time style transfer with simplified networks running on a high-end desktop GPU. However, achieving similar results on a full HD image on a mobile device would require tens of seconds. Elad and Milanfar's~\cite{elad2017style} approach takes multiple minutes to run on device.

Style transfer has several other drawbacks. First, as we will show in the results, output quality depends directly on the reference image. While often considered an advantage, having a specific template can generate inconsistent and undesirable results for different inputs (e.g., very bright/dark images). Second, current style transfer approaches do not provide sufficient aesthetic control for creating storyboards. Gatys et. al.~\cite{gatys2016controlling} extended their own work to introduce control over spatial location, color information, and scale. However, this does not allow the fine tuning required to design new stylizations. In this context, Barnes et al.~\cite{barnes2015patchtable} presented a method to efficiently query patches within a large dataset and replace each patch of the target image with one from the reference image. This approach does not allow control over color, line weight, or other aspects that characterize the hand drawn aesthetic of storyboards and comics.
\end{itemize}

Note that our method does not compete against these approaches, each of the techniques above can be incorporated into our system as a new block that further enriches expressiveness.

%
\section{Overview}\label{sec:overview}
Our system takes as input a set of images such as an album or a video sequence and converts them into one or more stylized storyboards. The aim is to create a visually pleasing graphic narrative that shows in a glance the input. Rather than creating a single unique summarization, our approach makes it easy to explore countless alternatives and quick to choose preferred variations in images, layout, and stylization.
Figure~\ref{fig:pipeline} describes the system pipeline:
\begin{enumerate}
  \item \textbf{Input:} The input of our system can be any video (e.g., a personal video sequence that last few tens of seconds) or any set of images (e.g., a personal photo album of a trip).
  \item \textbf{Frame Selection:} From the given input frames, we select a subset of frames that we will use to create the storyboard (Sec. ~\ref{sec:image_selection}).
  \item \textbf{Layout:} We choose a layout to hold the selected frames (Sec.~\ref{sec:layout}).
  \item \textbf{Framing:} We crop and zoom the selected frames to fit the layout (Sec. ~\ref{sec:framing}).
  \item \textbf{Stylization:} We stylize the resulting frame to give it a storyboard-like appearance (Sec.~\ref{sec:stylization}).
\end{enumerate}

\section{Image Selection}\label{sec:image_selection}

In this section, we describe how we select a subset of individual video frames or photos that will be used to create the final storyboard. Our approach needs to work for both videos and images and still run at interactive rates. Keyframe approaches and deep learning summarization proved too computationally expensive, so we explored simpler approaches.

For videos, we found that uniform frame sampling yields a representative set of input images. For image albums, we use all the available images as initial input set. In both cases, we remove duplicates and near-duplicates. 

\subsection{Duplicates}

Albums and videos contain visually similar scenes that might result in repetitive and redundant storyboards. Thus, we require a fast and reliable approach to detect near-duplicate images. In videos, most duplicates appear as successive frames; however, in photo albums it might not be the case, since it is common to collect images from different sources where the time stamp and viewpoint often differs.

One approach to finding duplicate files is to use hash functions, such as SHA-1 or MD5~\cite{rivest1992md5}. However, this limits the detection to exact duplicates. Because we are interested in near duplication, we use a modified version of perceptual hashing~\cite{nandisha2016piracy}. The process is as follows:
\begin{enumerate}
\item The image is converted to grayscale. 
\item The image is downscaled to $9\times8$ pixels. We downscale it recursively by $2$ until it reaches the closest scale factor and then apply a final spatial downscaling filter.
\item Gradient is computed vertically on odd rows and horizontally on even rows.
\item We create a 64-bit fingerprint according to the sign of the gradient, vertically in the odd rows and horizontally in the even rows.
\end{enumerate}

This \texttt{uint\_64} can be used as a fingerprint of each image and compared with other images using the Hamming distance, i.e., to compare two images we just count the number of different bits of their fingerprints. This approach, since it aggregates many pixels into one, is very robust to noise, color change, and small changes in camera position. Moreover, the distance can give an estimation of change between images.

\subsection{Detect Blurry Images}

Given a set of near-duplicate images, we select the sharpest image.

\noindent There are many methods in the literature for detecting blurry images. Ayd{\i}n et al.~\cite{aydin2015automated} use the residuals of iterative bilateral filters to compute a sharpness metric of the image. Talebi and Milanfar~\cite{2017arXiv170905424T} evaluate image quality using a CNN trained on images with different levels of distortion. However, these methods are too computationally expensive.

After analyzing different alternatives, we found the gradient of the image is the best indicator of sharpness. We compute a subsampled version of the gradient of the image as:
%
\begin{equation}
hist(i)=\#\{(x,y) \in \Omega: \lfloor \nabla(x,y) \rfloor = i\}
\end{equation}
where $\#$ is the cardinality, $\Omega$ is the subsample domain of the image (in our case, we subsample by a factor of $3$ in each direction), and $\nabla$ is the gradient (in our case, computed using central differences).

We then compute the \textit{sharpness metric} as the inverse of the \textit{coefficient of variation} of the histogram:
\begin{equation}
sharpness=\frac{\overline{hist}}{\sqrt{\overline{hist}_{2}}}
\end{equation}
where $\overline{hist}=
\frac{1}{\#\Omega}\sum_{i}{hist(i)}$, i.e., the mean value of the histogram and $\sqrt{\overline{hist}_{2}}=\frac{\sum_{i}{(hist(i)-\overline{hist})^2}}{\#\Omega}$, i.e., its standard deviation. The $\overline{hist}$ is already a good metric for sharpness between near-duplicates; we divide by the standard deviation to normalize the $sharpness$ measure so it is more robust to different content.

\section{Layout}\label{sec:layout}

Because we aim to create many alternative storyboards (in contrast to a unique optimal solution), we  initially believed automatic layout generation would provide the best approach. We based our initial efforts on the idea of parcel subdivision in urban planning~\cite{vanegas2012inverse}, i.e., given a city block (in this case the whole storyboard), how it can be divided into parcels (in this case panels).

\begin{figure}[tb]
\centering
\includegraphics[width=\linewidth]{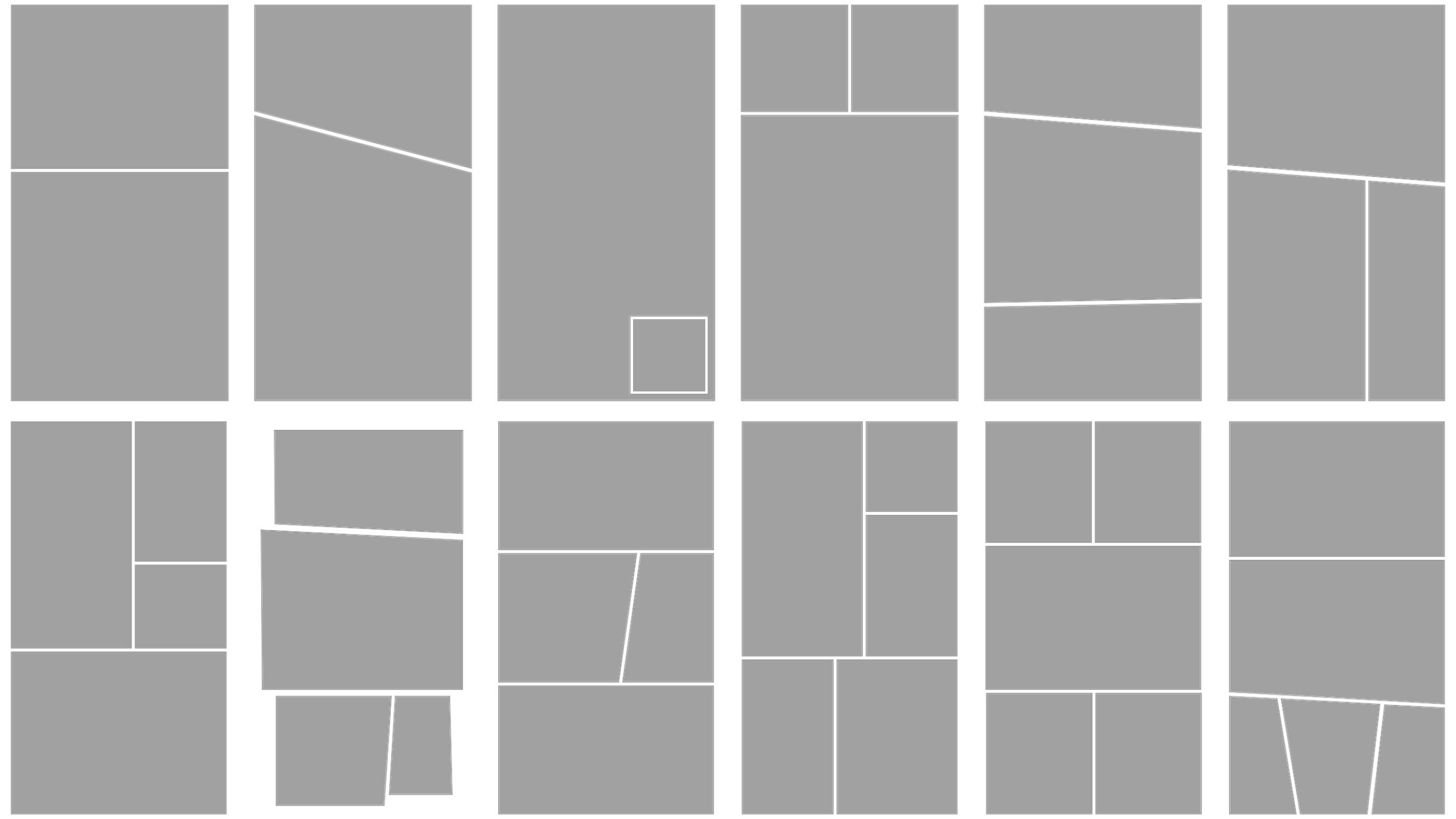}
\caption{\label{fig:fig_comic_template}Example set of layout templates.}
\end{figure}

This process yielded feasible but aesthetically unsatisfying layouts, so we opted instead to have designers manually create a set of over 70 predefined layouts. Some examples can be seen in Fig.~\ref{fig:fig_comic_template}.

Having a predefined set of layouts allows us to ensure consistently visually pleasing and non-repetitive results. Moreover, since we know ahead of time the position of all the panels, we can optionally merge content across adjacent panels while preserving their frames, a technique commonly employed by comic artists for dramatic effect.

\section{Framing: Cropping and Zoom}\label{sec:framing}

\begin{figure*}[tbp]
\centering
\includegraphics[trim={1.1cm 0.5cm 1.2cm 7.5cm},clip,width=\linewidth]{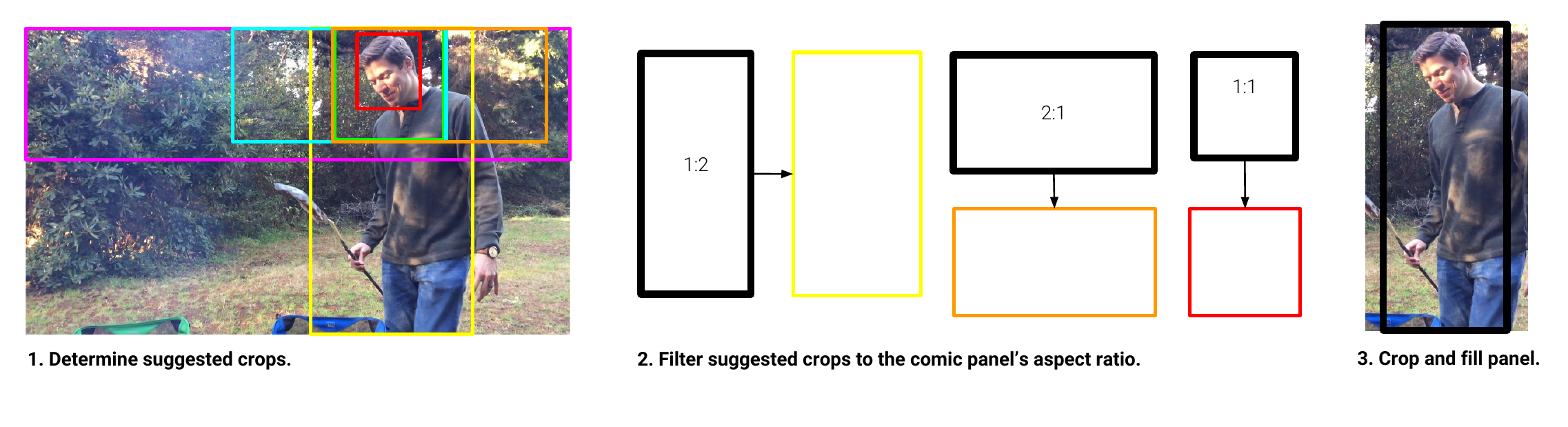}
\caption{\label{fig:crop}Framing: Cropping and Zoom.}
\end{figure*}

In this section, we explain how we frame images into layout panels. We use cropping and zooming to conform input images to fit specific panels. This allows us to explore more compositions than optimizing the layout to the selected frames -- especially and find more visually interesting crops when the image lacks recognizable objects (e.g., when the image is just a landscape).


Most summarization comic approaches rely on detecting faces to select the frames. We noticed that for our goal, despite faces being an essential part of videos and albums, just detecting faces would produce repetitive panels, especially in shorter videos. Basing frame selection on faces does not generalize to many other common video subjects such as places, vehicles, or pets. Thus, we use two CNNs: one trained to detect faces and another trained to detect objects. Since we wanted to be very fast and to be able to run on device, we use two \textit{single shot multibox detector} (SSD) presented by Liu et al.~\cite{liu2016ssd}. 

The SSD returns an approximate bounding box for detected objects, and from this, we generate a set of suggested crops (Fig.~\ref{fig:crop}.1). These crops are expansions of the detected bounding box (in red) with the following rules: extend 20 pixels, expand the crop bounding box's height and/or width by $10\%$ or $50\%$, or extend the bounding box to the image's full height or width. Once we have the expanded bounding boxes, we find the best match for a given target panel's aspect ratio (e.g., 1:2, 2:1, or 1:1) (Fig.~\ref{fig:crop}.2). Finally, we crop and zoom the input image to fill the target panel (Fig.~\ref{fig:crop}.3). We repeat this process until all the panels are populated.

\section{Stylization}\label{sec:stylization}

Stylization is the process of altering an image's appearance to make it look like another image (i.e., style transfer) or to make it more interesting or artistic, through filtering. As we show in the results (Sec.~\ref{sec:results}), we discarded style transfer given its complexity and since it does not allow the required control (efforts such as~\cite{gatys2016controlling} just allows control within the same style). Therefore, we focus on stylization through filtering.

In this section, we describe the interactive tool we created for designing styles by combining filter blocks.

\subsection{Interactive Style Design}

While most stylization and abstraction works (Sec.~\ref{subsec:stylization}) focus on creating one filter or a fixed set of filters to achieve a stylization, our goal is to create a flexible tool that allows anyone to design stylization filters, regardless of technical ability.

\begin{figure}[tb]
\centering
\includegraphics[width=\linewidth]{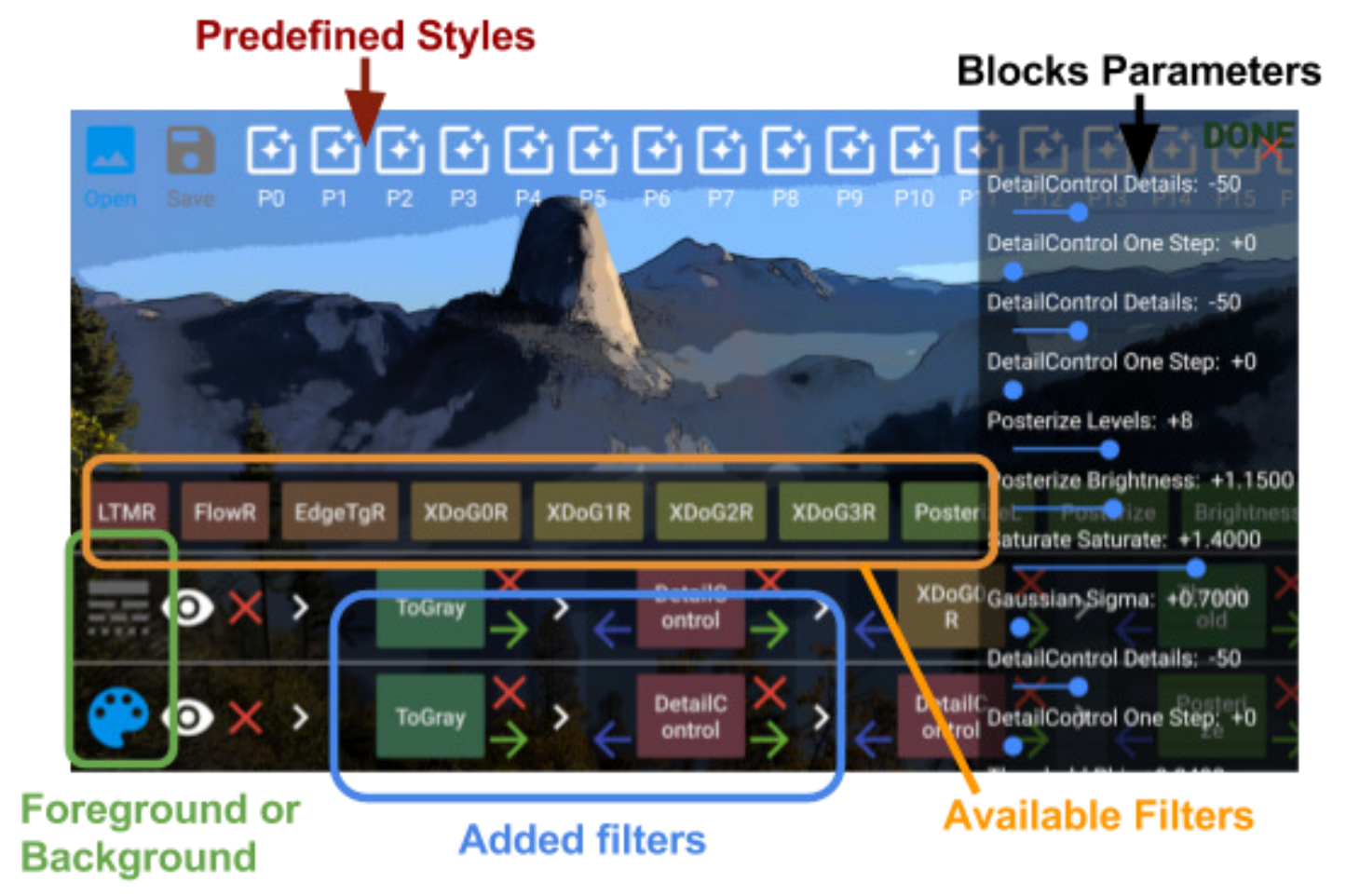}
\caption{\label{fig:fig_style_design_app}Style Design App.}
\end{figure}

The final result of our framework is as shown in Fig.~\ref{fig:fig_style_design_app}. The main components are:
\begin{enumerate}
\item A wide range of filter blocks. (See Sec.~\ref{subsec:filters})
\item Parameters for each filter that can be controlled by sliders that appear on the right side of the screen.
\item Two layers: a black and white foreground layer used as alpha channel to display lines and contours (e.g., using XDoG or Sobel filtering); and a background layer for color stylization. This separation provides added design flexibility.
\item A visual flow diagram of filter blocks which allow any block filter to be added, moved, reordered, removed, and tuned at any time. This is the essence of the system's interaction design.
\end{enumerate}

\subsection{Filter blocks}\label{subsec:filters}

\begin {figure*}[hbt]
\centering
\begin{adjustbox}{width=\textwidth}
\begin{tikzpicture}
\node [inner sep=0pt,above right] (img) at (0,0){\includegraphics[width=\textwidth]{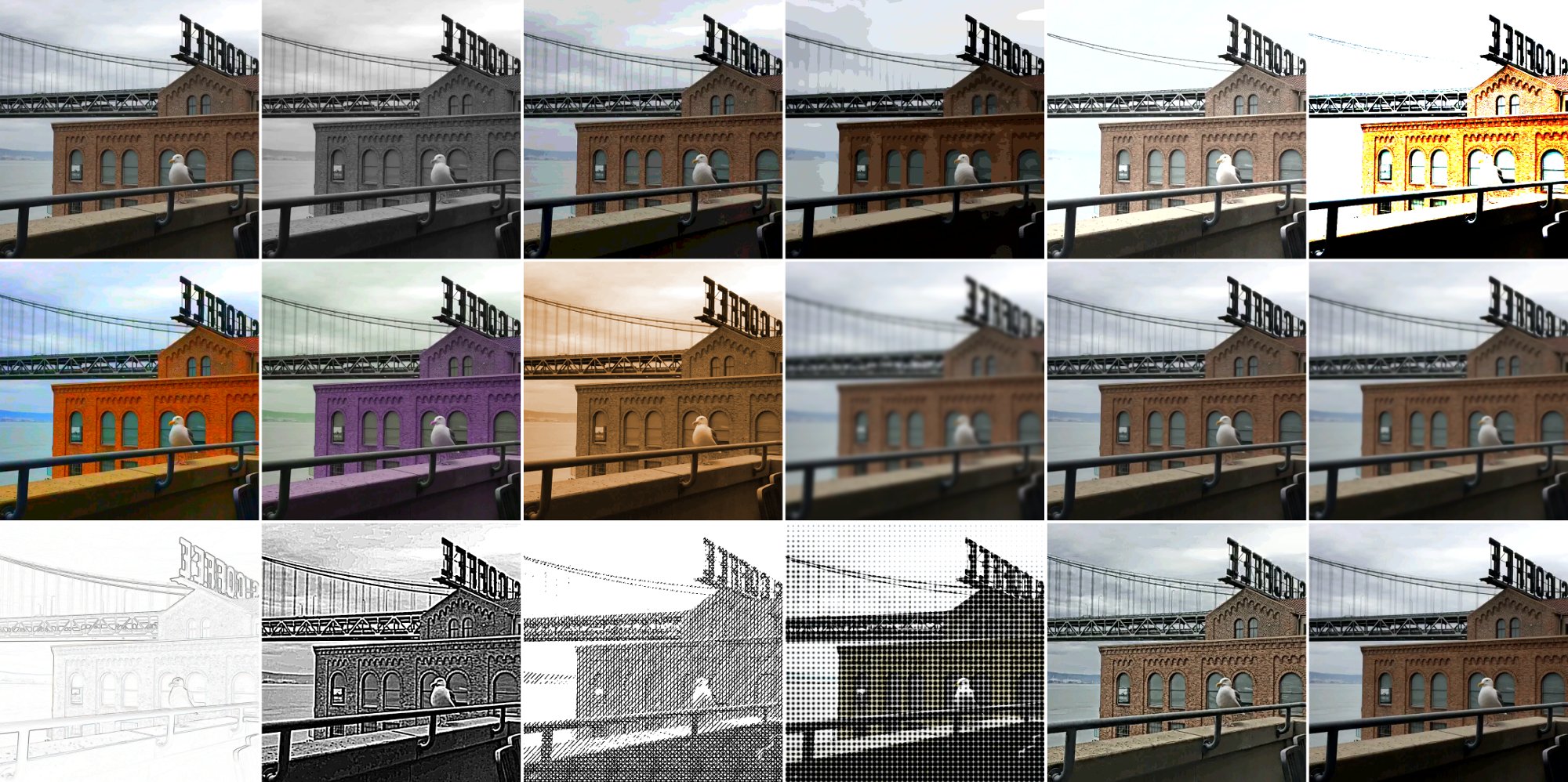}};
\begin{scope}[x=(img.south east),y=(img.north west)]
\foreach [count=\i] \y in {2,...,0}{
  \foreach [count=\j] \x in {0,...,5}{
     \draw [fill=white,white] (\x/6+0.0004,\y/3) rectangle (\x/6+0.02,\y/3+0.04);
     \draw (\x/6+0.01,\y/3+0.02) node[black] {\alphalph{\j+6*(\i-1)}};
          }}
\end{scope}
\end{tikzpicture}
\end{adjustbox}
\caption{\label{fig:all_filters}Filters: a) Input, b) To GrayScale, c) Posterization, d) Luma Posterization, e) Brightness, f) Soft Threshold, g) Saturation, h) Hue, i) Colorize, j) Gaussian Smoothing,  k) ETF, l) TVF, m) Sobel, n) XDoG, o) Pattern Filling, p) Halftone, q) Image Detail Control, and r) Linear Equalization. See text for details.}
\end{figure*}

We implemented three kinds of blocks: pixel operations, advanced filters, and histogram modification filters. Fig.~\ref{fig:all_filters} shows the effect of each individual block filter.


\noindent We included the following list of pixel operations:

\begin{itemize}
\item \textbf{To GrayScale:} Converts the image into grayscale/luma and saves the chromatic channels ($UV$). This block is useful for applying any other block filter to just the luma channel (for aesthetics or performance reasons). (Fig.~\ref{fig:all_filters}.b).
\item \textbf{To Color:} Uses the current luma and converts the image back to $RGB$ using the $UV$ previously saved by the \textit{To GrayScale} block.
\item \textbf{Posterization:} Discretizes continuous image colors (e.g., 255 levels) to regions of fewer tones (e.g., $levels = 10$). (Fig.~\ref{fig:all_filters}.c)
\item \textbf{Luma Posterization:} Posterizes the image in the luma channel by converting the image to grayscale, applying \textit{Posterization}, and converting it back to color. (Fig.~\ref{fig:all_filters}.d).
\item \textbf{Brightness:} Multiplies the luma channel by a user-selected $brightness$ constant, clipping the output values. (Fig.~\ref{fig:all_filters}.e).
\item \textbf{Soft Threshold~\cite{winnemoller2012xdog}:} Performs the following expression for each pixel $out(x,y) = 255 * (1 + \tanh(\min(0, \phi * (in(x,y) - \epsilon))))$ where $\phi$ determines the slope and $\epsilon$ the cut-off. For grayscale images, this block behaves like a binary cut-off that preserves smooth transitions. For color images, it simplifies each $RGB$ channel into two levels. (Fig.~\ref{fig:all_filters}.f).
\item \textbf{Saturation:} Makes the colors more vivid (saturates) or more muted (desaturates) by adding or subtracting in $RGB$ the grayscale image tuned by a parameter. (Fig.~\ref{fig:all_filters}.g).
\item \textbf{Hue:} Performs a color rotation in $UV$ space and adds a bias in $RGB$. This block is useful for changing the image's tint. (Fig.~\ref{fig:all_filters}.h).
\item \textbf{Colorize:} Convert to monochrome using an $HSL$ palette transformation. (Fig.~\ref{fig:all_filters}.i).
\end{itemize}

\noindent We also included more advance filters:

\begin{itemize}
\item \textbf{Gaussian Smoothing:} Blurs and removes details and noise from the image. The standard deviation ($\sigma$) is controlled by a parameter. (Fig.~\ref{fig:all_filters}.j).
\item \textbf{Edge Tangent Flow (ETF)~\cite{kang2007coherent}:} Creates an impressionistic oil painting effect. This method uses a kernel-based nonlinear smoothing of vector field inspired by bilateral filtering. It is implemented as described at~cite{2017arXiv171110700G}. (Fig.~\ref{fig:all_filters}.k).
\item \textbf{Total Variation Flow (TVF)~\cite{louchet2011total}:} Makes image piecewise constant with an anisotropic diffusion filter. It is implemented as described at~\cite{2017arXiv171110700G}. (Fig.~\ref{fig:all_filters}.l)
\item \textbf{Sobel Filter~\cite{sobel1990isotropic}:} Fast edge-detection filter. (Fig.~\ref{fig:all_filters}.m).
\item \textbf{XDoG~\cite{winnemoller2012xdog}:} Uses \textit{Difference-of-Gaussians} to find the edges of the image. In our implementation, the user can control the variance of the main Gaussian ($\sigma$) and the multiplier ($p$). (Fig.~\ref{fig:all_filters}.n)
\item \textbf{Size:} Upscales or downscales the image using a user-selected scale parameter. This block can help to speed up computation (computing other blocks in a lower resolution) and alter the behavior of other scale-dependent filters.
\item \textbf{Pattern Filling Filter:} Uses \textit{Luma Posterization} to discretize the image into a set of levels, then, each pixel is replaced by a texture depending on its level. This block is useful for creating cross-hatching patterns. (Fig.~\ref{fig:all_filters}.o).
\item \textbf{Halftone:} Replaces colors with a set of dots that vary in size and color. This style mimics the behavior of the four-color printing process traditionally used to produce comics. (Fig.~\ref{fig:all_filters}.p)
\item \textbf{Image Detail Control:} Inspired by~\cite{talebi2016fast}, controls the details of the image by adding the residual of the image to its bilateral version multiplied by a $\delta$. $\delta < 0$ smooths the image while $\delta>0$ adds details. (Fig.~\ref{fig:all_filters}.q).
\end{itemize}

\noindent Finally, we included two blocks to modify the histogram:
\begin{itemize}
\item \textbf{Linear Histogram Equalization:} Equalizes the luma channel expanding the $p_{l}$ to zero and the $p_{h}$ to 255. Normally we choose $l=5$ and $h=95$ such that the $5\%$ percentile is moved to zero and the $95\%$ is moved to 255, thereby increasing the image's dynamic range. (Fig.~\ref{fig:all_filters}.r).
\item \textbf{Histogram Minimum Dynamic Range:} Computes the percentile $5\%$ and $95\%$ on the luma histogram and expands (if necessary) the dynamic range to match the user parameter range $DR$.
\end{itemize}

These two filters are usually placed as the first filter in the pipeline to force the image to have a proper dynamic range to obtain a satisfactory result (e.g., XDoG on a hazy or too bright or dark image would result in an almost entire white output image).

\section{Results}\label{sec:results}

In this section, we present the following results: a set of styles produced by designers, a set of styles generated procedurally, an example of video converted to storyboards, and finally a few examples of style transfer.

\begin{figure}[tb]
\centering
\includegraphics[width=\linewidth]{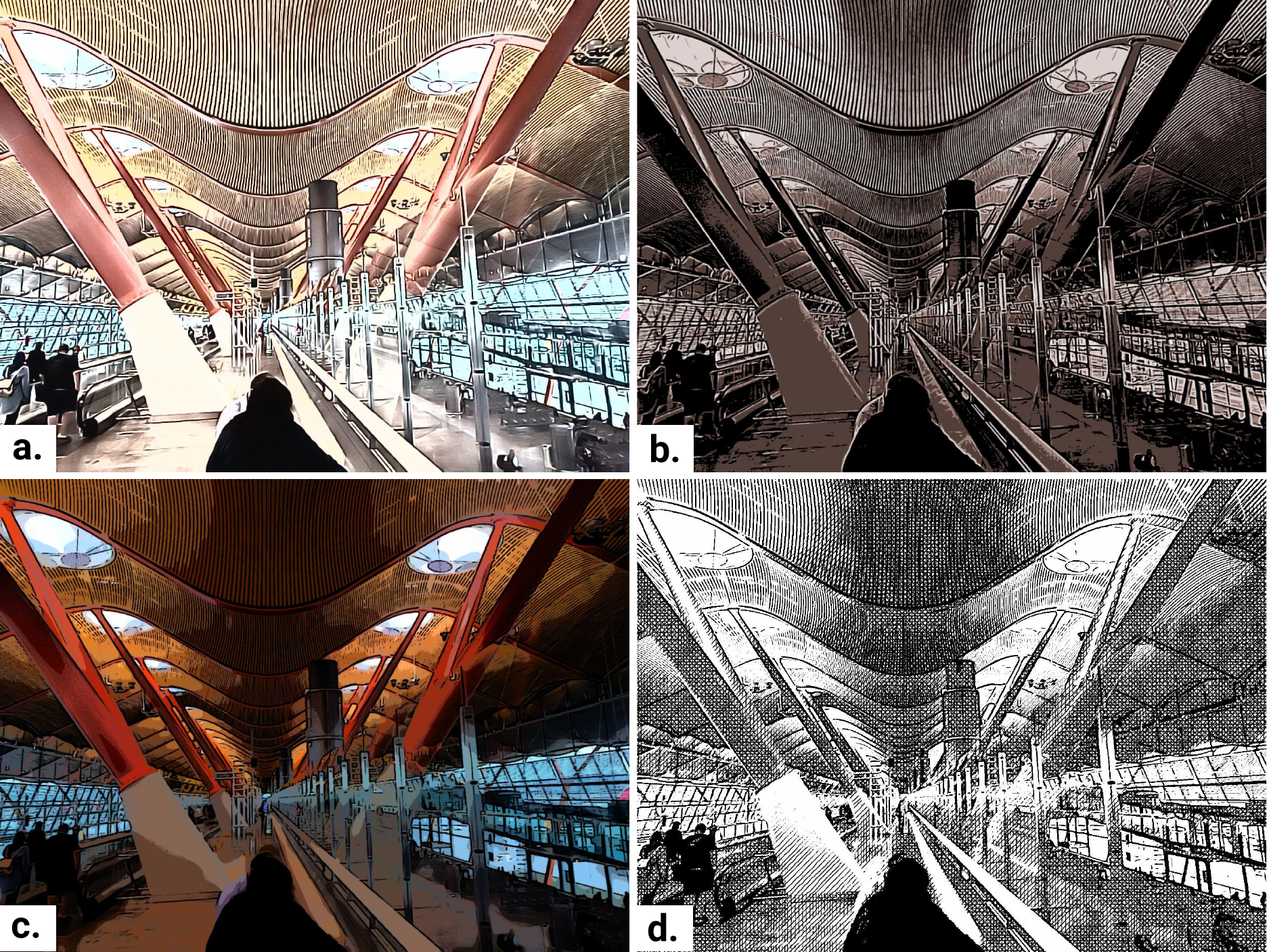}
\caption{\label{fig:fig_styles_no_timings}Four styles generated by our design tool.}
\end{figure}

\noindent\textbf{Designed Styles and Timings:}\\
We conducted several style design sessions with designers who generated dozens of styles that were suitable for storyboard-like stylization. Fig.~\ref{fig:fig_styles_no_timings} shows four examples of these styles and Table~\ref{tab:computation_time} lists the time for computing the filter blocks for each style. The first filter uses XDoG and Soft Threshold to create the main effect. The second filter, which produces the outlines, uses XDoG with Threshold (strong lines) and, for the background, uses Pattern. The third filter uses a smoothed version of the image to compute the outlines, and the color is achieved by smoothing the image and applying posterization. The final filter uses ETF, Posterize, and Colorize. The average time to compute a Full HD ($1920\times1080$) image on desktop (Xeon E5-1650v3) is $72.45\units{ms}$ and on a Nexus 6P is $291.625\units{ms}$. The code is optimized using the Halide programming language~\cite{ragan2012decoupling}. Halide code decouples algorithm implementation from scheduling. This allows us to create optimized code that utilizes native vector instructions while parallelizing over multiple CPU cores.


\begin{table}[tb]
\caption{\label{tab:computation_time}Style computation time of Fig.~\ref{fig:fig_styles_no_timings} (in ms) for \textit{`Desktop'} (Xeon E5-1650v3) and \textit{`Device'} (Nexus 6P). For each example, we list the background filters first and then the foreground filters (if any) separated by a line.}

\begin{minipage}{.5\linewidth}
\small
\begin{tabular}{l|cc} 

\textbf{Fig.~\ref{fig:fig_styles_no_timings}.a}  & Desktop & Device \\ \hline\hline
ToGray    & 2.0     & 15.8   \\
XDoG      & 27.9    & 107.6  \\ \hline
Thresh. & 0.4     & 1.9    \\
ETF       & 18.3    & 72.4   \\
ToRgb     & 4.3     & 15.1   \\ \hline\hline
Total & 52.9\scriptsize{ms} & 212.8\scriptsize{ms}
\end{tabular}

\bigskip

\begin{tabular}{l|cc}
\textbf{Fig.~\ref{fig:fig_styles_no_timings}.c}     & Desktop & Device \\ \hline\hline
Detail C.    & 27.0    & 125.4  \\
ToGray         & 1.2    & 4.8    \\
Posteriz.      & 0.3    & 2.0    \\
ToRgb          & 3.8     & 15.3   \\
Saturate       & 3.4     & 11.0   \\ \hline
ToGray         & 1.6     & 4.7    \\
Detail C.    & 17.8    & 92.2   \\
XDoG           & 22.5    & 93.9   \\
Threshold      & 0.5     & 2.3    \\
TVF            & 20.6    & 85.3   \\ \hline\hline
Total & 98.7\scriptsize{ms} & 436.9\scriptsize{ms}
\end{tabular}

\end{minipage}
\begin{minipage}{.5\linewidth}
\small
\begin{tabular}{l|cc}
\textbf{Fig.~\ref{fig:fig_styles_no_timings}.b}    & Desktop & Device \\ \hline\hline
ToGray         & 1.3     & 5.7    \\
ETF            & 17.6    & 77.1   \\
XDoG           & 23.5    & 92.1   \\
Posteriz.      & 0.5     & 1.9    \\
Colorize       & 7.5     & 12.0   \\
Detail C.    & 38.1    & 132.8  \\ \hline\hline
Total & 88.5\scriptsize{ms} & 321.6\scriptsize{ms}
\end{tabular}

\bigskip

\begin{tabular}{l|cc}
\textbf{Fig.~\ref{fig:fig_styles_no_timings}.d}     & Desktop & Device \\ \hline\hline
ToGray    & 1.4     & 4.8    \\
Threh.  & 0.3     & 2.0    \\
Pattern   & 3.7     & 10.3   \\
TVF       & 20.7    & 77.2   \\ \hline
XDog      & 23.1    & 89.9   \\
Thresh. & 0.5     & 2.0    \\ \hline\hline
Total & 49.7\scriptsize{ms} & 186.2\scriptsize{ms}
\end{tabular}

\end{minipage}
\end{table}

\noindent\textbf{Video to Storyboard:}\\
\begin{figure*}[tbh]
\centering
\includegraphics[width=\linewidth]{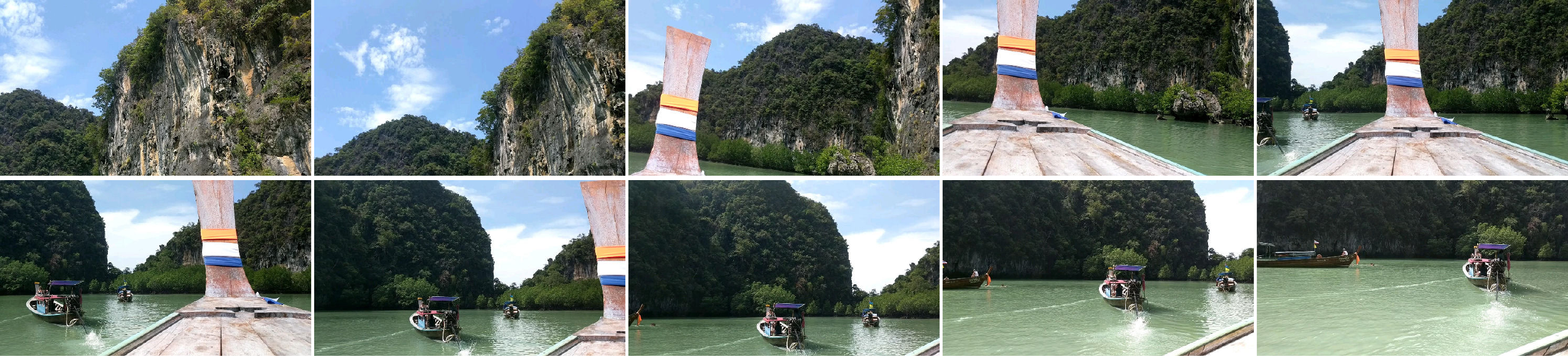}
\caption{\label{fig:video_seq}All the keyframes extracted from a 16-second input video. This is the input for the results pictured in Fig.~\ref{fig:video_to_comic}.}.
\end{figure*}
Fig.~\ref{fig:video_to_comic} shows one example of our system converting a previously unedited video (Fig.~\ref{fig:video_seq}) into storyboards. Our system randomly selects the layout, stylization, and creates the variations. Note that the pictured examples were not cherry-picked; they are representative of the system's normal behavior.
\\ \\ 

\begin{figure*}[tbh]
\centering
\includegraphics[width=\linewidth]{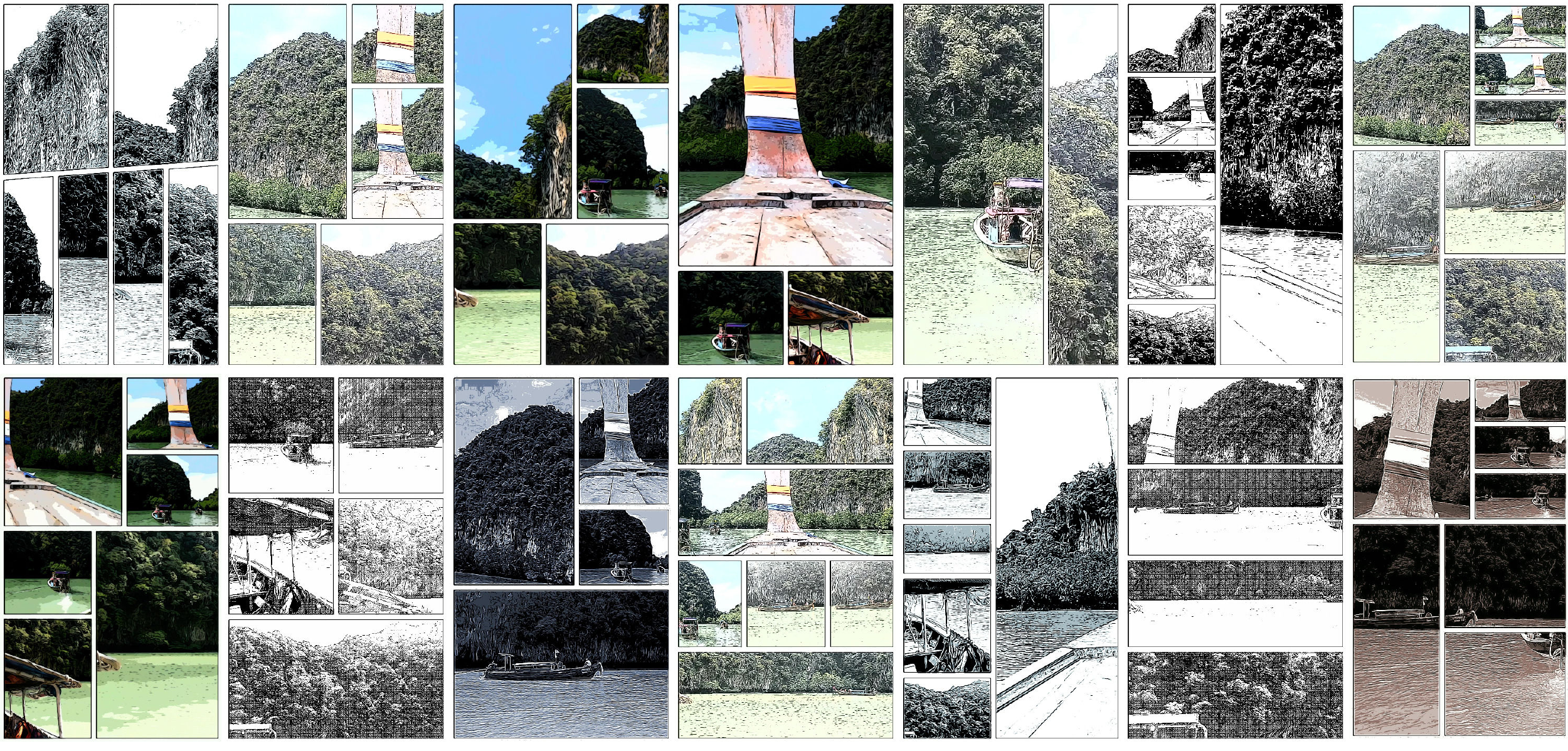}
\caption{\label{fig:video_to_comic}Example output of our system for the input images shown in Fig.~\ref{fig:video_seq}. The images have not been cherry-picked; they are the first 14 layouts generated by our system for this input.}
\end{figure*}

\noindent\textbf{Procedural Styles:}\\
Procedural modeling is a set of techniques that is able to generate hundreds or thousands of examples using a limited set of rules. In the computer graphics community, the technique has been used to generate buildings, cities, trees, and more complex models (e.g.,~\cite{muller2006procedural,nishida2016interactive}). Based on this idea, we call \textit{procedural styles} those styles created randomly using a set of simple rules. This allow us to create thousands of styles and explore  Given that our system can assemble any sequence of filters (just restricted by the number of input channels for some filters), we decided to randomly generate several thousands of styles and choose the best ones.

\begin{figure}[tb]
\centering
\includegraphics[width=\linewidth]{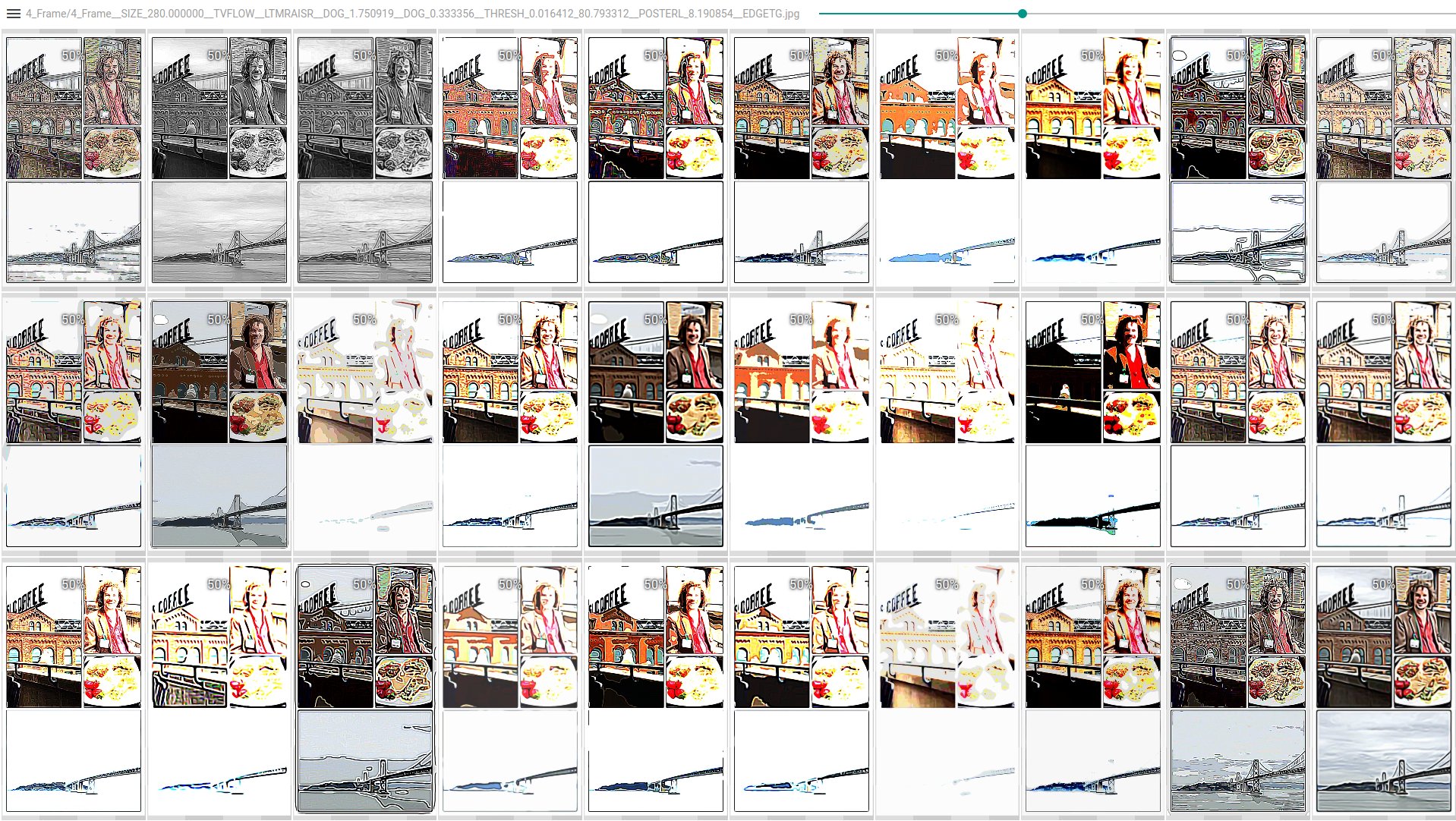}
\caption{\label{fig:exploration_tool}Visualization tool to explore alternatives. Most alternatives (as seen in the figure) are not very interesting. However, the visualization tool makes it fast to explore and identify promising alternatives.}
\end{figure}

\begin{figure}[tb]
\centering
\includegraphics[width=\linewidth]{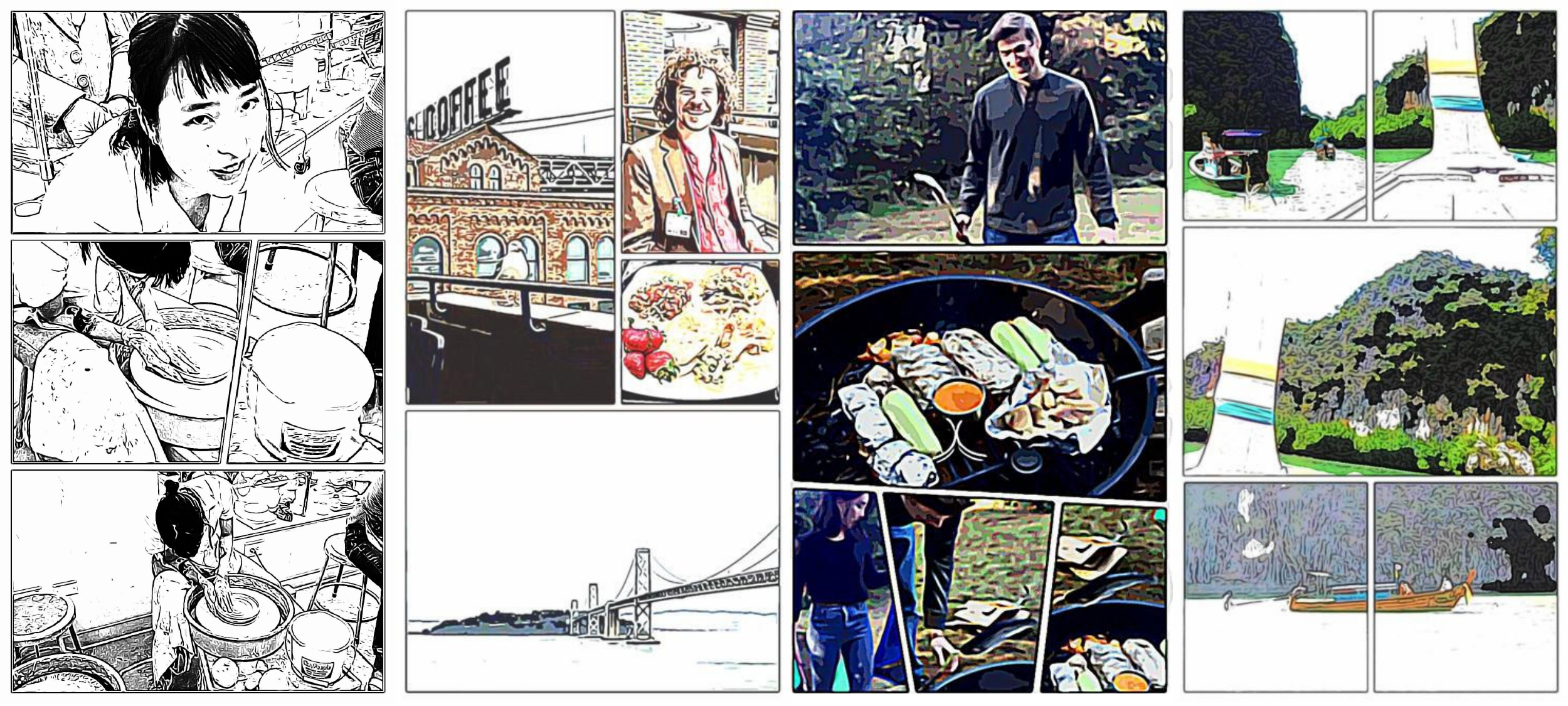}
\caption{\label{fig:just_random}Top-scored randomly selected styles.}
\end{figure}

The simple set of rules for this experiment was to add a random number of filters between 4 and 9 chosen randomly with random input parameters: XDoG ($\sigma \in [0.5, 8.0]$ and $p \in [1, 40]$), TVF, Soft Threshold ($\phi \in [0.013, 0.059]$ and $\epsilon \in [50, 110]$), Detail Control ($\delta \in [-100, 60]$), Luma Posterization ($level \in [5, 12]$), Saturation ($saturation \in [1.5, 2.2]$), Size ($size \in [100, 300]$, and To GrayScale ($20\%$ probability). We also enforce that XDoG and TVF are the only filters that can be added more than once (duplicating the rest of filters has the same effect as selecting different parameters).

We developed a web-based visualization tool to quickly review the randomly generated alternatives (Fig.~\ref{fig:exploration_tool}). In order to automate this process, we used a CNN approach~\cite{2017arXiv170905424T} that evaluates the aesthetic quality of an image. Using this approach, we found the styles in Fig.~\ref{fig:just_random}.
\\ 

\begin{figure}[tb]
\centering
\includegraphics[width=\linewidth]{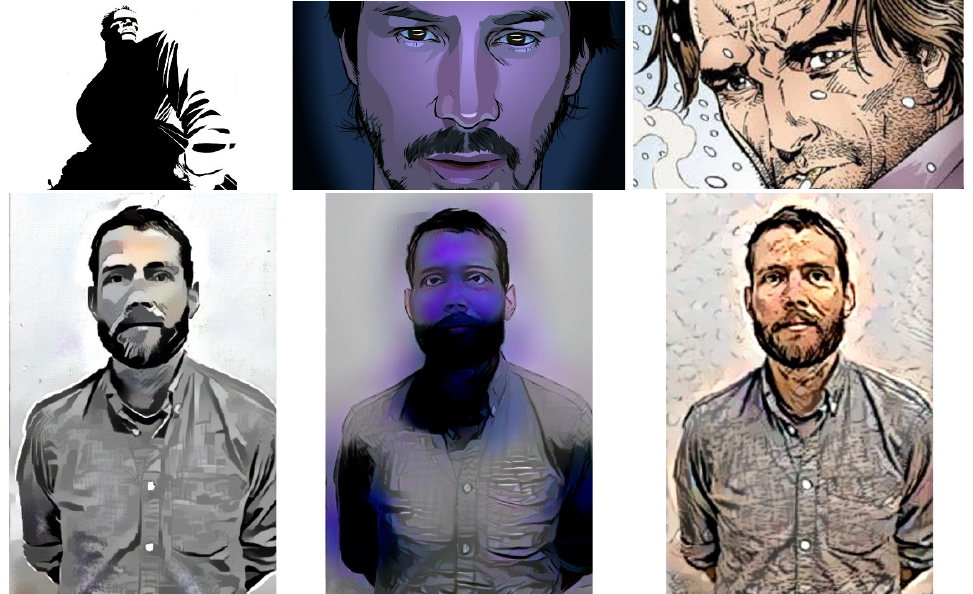}
\caption{\label{fig:fig_cnn}Examples of style transfer done using~\cite{gatys2016image}.}
\end{figure}

\noindent\textbf{Style Transfer:}\\
One interesting alternative to using stylization through filtering, as described before, it is to use style transfer. We conducted a thorough analysis and could not find a template image that would work for all test inputs. Fig.~\ref{fig:fig_cnn} shows several examples we explored. We found the result was strongly dependent on the context, i.e., consistent results would occur only when the template and the input had similar properties (e.g., similar content with different styles). The high complexity of this algorithm is also significant. Gatys et al.~\cite{gatys2016image} report it takes up to an hour on a graphics card to compute a $512\times512$ image. Extensions of this work that focus on speed, e.g.,~\cite{johnson2016perceptual}, achieve real-time performance using a high-end desktop GPU, however, it would require dozens of seconds to run on a Full HD image on a mobile device, which makes these approaches impractical for our purposes.

\section{Conclusions and Future Work}\label{sec:conclusions}

This paper described a system for automatically converting an album or a video into hundreds of storyboards. We described the process through its main steps: image selection, framing, and stylization. The end result takes a set of images and converts them into an interesting, visually pleasing graphic representation.

In this work, we also presented the first interactive framework for designing filter-based stylization. This approach boosts the creativity by allowing the designer to tune, modify, and play with the filters directly. In parallel to this manual design, we presented a procedural stylization that follows a list of simple rules to create hundreds of styles automatically. These styles can be selected through manual visualization or using a previously trained aesthetics quality assessment neural network. 

As future work, we may pursue several items. We would like to explore using the audio from the input video, possibly by detecting and labelling the audio to add sound effects to the resulting storyboard~\cite{gemmeke2017audio}. As an example, applause in the video might be converted into a stylized sticker with the words "clap clap." Also, we plan to use \textit{scene descriptor}~\cite{karpathy2015deep} to caption the storyboard. As an alternative to our stylization, we would also like to explore how we could extend CNN-based approaches to fit our context.


\bibliographystyle{eg-alpha}
\newcommand{\etalchar}[1]{$^{#1}$}

\newpage

\end{document}